\documentclass[fleqn,usenatbib]{mnras}
\usepackage{newtxtext,newtxmath}
\usepackage[T1]{fontenc}
\usepackage{ae,aecompl}

\usepackage{graphicx} % Including figure files
\usepackage{amsmath,amssymb,times}
\usepackage{epstopdf}
\usepackage[usenames,dvipsnames]{color}
\usepackage{float}
\usepackage[export]{adjustbox}
\usepackage{ulem}
\usepackage{hyperref}
\usepackage{pdflscape}
\usepackage[utf8x]{inputenc}

\hypersetup{
            colorlinks = true, %Colours links instead of ugly boxes
            urlcolor = blue, %Colour for external hyperlinks
            filecolor = black,
            runcolor = blue,
            linkcolor = red, %Colour of internal links
            citecolor = blue %Colour of citations
            }
\makeatletter
\newcommand{\kms}{km\,s$^{-1}$}
\newcommand{\NII}{N\,{\scriptsize II}}
\newcommand{\OIII}{O\,{\scriptsize III}}

\newcommand{\HI}{H\,{\scriptsize I}}

\newcommand{\nad}{Na\,{\scriptsize D}}
\newcommand{\cii}{C\,{\scriptsize II}}
\newcommand{\Hii}{H\,{\scriptsize II}}
\newcommand{\MgII}{Mg\,{\scriptsize II}}
\newcommand{\FeII}{Fe\,{\scriptsize II}}
\makeatother

\makeatletter
\def\@to{to}
\makeatother

\title[The Multiphase Nature of Outflows]{Observational Constraints on the Multiphase Nature of Outflows Using Large Spectroscopic Surveys at $z\sim$0}

\author[Roberts-Borsani, G.~W.]{G.~W. Roberts-Borsani$^{1,2}$\thanks{E-mail: guidorb@astro.ucla.edu}
\\
$^{1}$Department of Physics and Astronomy, University of California, Los Angeles, 430 Portola Plaza, Los Angeles, CA 90095, USA \\
$^{2}$Department of Physics and Astronomy, University College London, Gower Street, London WC1E 6BT, UK}

\date{Accepted XXX. Received YYY; in original form ZZZ}
\pubyear{2020}

\begin{document}
\label{firstpage}
\pagerange{\pageref{firstpage}--\pageref{lastpage}}
\maketitle

\begin{abstract}
Mass outflow rates and loading factors are typically used to infer the quenching potential of galactic-scale outflows. However, these generally rely on observations of a single gas phase which can severely underestimate the total ejected gas mass. To address this, we use observations of high mass ($\geqslant$10$^{10}$ M$_{\odot}$), normal star-forming galaxies at $z\sim$0 from the MaNGA, xCOLD GASS, xGASS and ALFALFA surveys and a stacking of \nad, H$\alpha$, CO(1-0) and \HI\ 21cm tracers with the aim of placing constraints on an average, total mass outflow rate and loading factor. We find detections of outflows in both neutral and ionised gas tracers, with no detections in stacks of molecular or atomic gas emission. Modelling of the outflow components reveals velocities of $|$v$_{\text{NaD}}|$=131\,\kms\ and $|$v$_{\text{H}\alpha}|$=439\,\kms\ and outflow rates of $\dot{M}_{\text{NaD}}$=7.55\,M$_{\odot}$yr$^{-1}$ and $\dot{M}_{\text{H}\alpha}$=0.10\,M$_{\odot}$yr$^{-1}$ for neutral and ionised gas, respectively. Assuming a molecular/atomic outflow velocity of 200\,\kms, we derive upper limits of $\dot{M}_{\text{CO}}<$19.43\,M$_{\odot}$yr$^{-1}$ and $\dot{M}_{\text{HI}}<$26.72\,M$_{\odot}$yr$^{-1}$ for the molecular and atomic gas, respectively. Combining the detections and upper limits, we find average total outflow rates of $\dot{M}_{\text{tot}}\lesssim$27\,M$_{\odot}$yr$^{-1}$ and a loading factor of $\eta_{\text{tot}}\lesssim$6.39, with molecular gas likely contributing $\lesssim$72\% of the total mass outflow rate, and neutral and ionised gas contributing $\sim$28\% and $<$1\%, respectively. Our results suggest that, to first order, a degree of quenching via ejective feedback could occur in normal galaxies when considering all gas phases, even in the absence of an AGN.
\end{abstract}


\section{Introduction}
The energy and momentum delivered by massive stars into the interstellar medium (ISM) through supernovae, strong stellar winds, cosmic rays and high energy photons - collectively known as ``feedback'' - can give rise to galactic-scale gas outflows, which serve as an integral component in the regulation of gas in and out of galaxies and their evolution as a whole \citep{heckman2000,veilleux05,rupke05b,lilly13}. Specifically, by ejecting large fractions of gas out of the galaxy disk, strong outflows (from extreme star formation or a supermassive black hole) provide one potential mechanism by which to halt the stellar mass growth of galaxies, through the quenching of the star formation in the disk. Such a process is considered vital towards reconciling the differences between the observed baryonic mass function and the mass function of $\Lambda$CDM-predicted dark matter halos \citep{bell03,somerville08,li09,bouche10}.

Studies of neutral, ionised and molecular gas outflows have made important progress towards constraining the properties of outflows and correlations with their host galaxies in recent years \citep{veilleux05, rupke05a, chen10, feruglio10, martin12, cicone14, rubin14, fiore17, rupke17, fluetsch19, rb19}, with the two primary purposes of such studies being to determine the prevalence of outflows in galaxies and active galactic nuclei (AGN) and their potential for quenching star formation. In constraining the prevalence of outflows, most progress has has been made via absorption- and emission-line studies of neutral and ionised gas in star-forming, (U)LIRG and AGN objects, both in the local Universe and extending out to $z\sim$2 \citep{rupke05b, chen10, newman12a, rubin14, cicone16, sarzi16, perna17, davies19, rb19}. Such studies, typically using \nad, \MgII\ or \FeII\ in absorption and H$\alpha$ or [\OIII] in emission, have generally found outflows to reside in massive (M$_{*}\gtrsim$10$^{10}$ M$_{\odot}$) galaxies either along or above the so-called galaxy ``main sequence'' (MS; \citealt{noeske07}), with detection rates increasing as a function of star formation or AGN activity \citep{heckman2000, chen10, newman12a}. These results also hold on kpc-scales, with IFU observations revealing prominent outflow signatures in regions of high $\Sigma_{\text{SFR}}$ ($\Sigma_{\text{SFR}}\gtrsim$0.01 M$_{\odot}$yr$^{-1}$kpc$^{-2}$) and AGN luminosity \\
\citep{ho-i16,lopezcoba19,rodriguezdelpino19,rb20}, and across redshift with significant detection rates found out to $z\sim$2 \citep{newman12a, genzel14, sugahara17, davies19}.

A clear consensus on the quenching potential of outflows, however, has yet to be established. Studies of cold, molecular gas tracers (typically CO(1-0)) in extreme starbursts and AGN reveal powerful outflows with velocities up to several 1000 km\,s$^{-1}$ and mass loss rates of 10-1000 M$_{\odot}$\,yr$^{-1}$ \citep{feruglio10, sturm11, cicone14, rupke17}, whilst observations of representative galaxies show significantly less powerful outflows with velocities and mass loss rates of only several 100 km\,s$^{-1}$ and a few M$_{\odot}$\,yr$^{-1}$ \citep{rupke05b, chen10, martin12, rb19}. Due to the faintness of outflow signatures and the scarcity of deep molecular and atomic gas observations for normal galaxies, the latter estimates have typically been made using rest-frame UV and optical tracers of neutral and ionised gas. The so-called ``mass loading factor'' ($\eta$), defined as the mass outflow rate per unit of star formation rate (SFR) and an important regulating component of current galaxy evolution models and simulations \citep{bouche10,oppenheimer10,lilly13}, is typically used to quantify the quenching potential of outflows and found to vary between unity and a few for representative star-forming galaxies \citep{martin12, rubin14, rb19}. Similarly for trends of outflow velocities and mass loss rates, only extreme starburst galaxies or AGN appear capable of significantly enhancing their mass loading factors to values of $\eta>>$1 that are suggestive of quenching \citep{rupke05b, cazzoli16, cicone14, fluetsch19}.

Despite progress in constraining the mass loading factor and quenching potential of outflows across galaxy populations, two glaring issues remain. The first is the tendency to constrain the parameter over relatively small and biased samples of extreme objects (major mergers, (U)LIRGs, starbursts and AGN) rather than statistical samples of normal galaxies, since the former produce more extreme outflows that are more easily observed by current telescope facilities. However, such objects are not representative of the general galaxy populations that likely transit from the main sequence of star-forming galaxies to the passive sequence. The second, and perhaps more challenging issue, is the highly uncertain comparison of mass loss rates and loading factors derived from single gas phase tracers. In addition to the fact these often require highly uncertain assumptions (e.g., outflow geometry and extent, gas ionisation corrections and depletion onto dust) to convert gas column densities to mass outflow rates \citep{chisholm16} - which can lead to order-of-magnitude uncertainties - the key concern is that neglecting other gas phases would severely underestimate the total gas mass ejected out of the galaxy and result in misleading and incomplete conclusions on the quenching potential of outflows \citep{ciconenature,chisholm18}. Although some efforts to constrain the multiphase nature of outflows have been made (e.g., \citealt{rupke13,tadhunter14,morganti16,rupke17,fluetsch19}), these have largely been exclusive to single or small samples of extreme or unusual objects and thus conclusions for normal star-forming galaxies lack or remain unclear.

As such, in order to obtain an unbiased view of outflow quenching in galaxies, observations of the multiphase nature in large samples of representative galaxies are crucial. With the above in mind, the next logical step - and indeed the aim of this short paper - is to place constraints on the prevalence and properties of multiphase outflows for statistical samples of representative galaxies, in order to gain a more thorough understanding of the relative fractions of gas ejected by outflows and whether these considerations have an impact on their quenching potential. In this work, therefore, we make use of stacking techniques of spectroscopic data sets from large surveys of normal galaxies from xCOLD GASS, xGASS, ALFALFA and MaNGA DR15 to place constraints on the multiphase nature of outflows. The paper is organised as follows. In \S2 we describe out sample selection from the surveys, based on results from previous studies. In \S3 we describe out spectroscopic stacking methods and validation of our methods and \S4 presents the results of our analysis. In \S5 we provide a discussion of our findings, with our summary and conclusions presented in \S6. Throughout this paper we assume \textit{H}$_{0}$=70 km\,s$^{-1}$\,Mpc$^{-1}$, $\Omega_{m}=$ 0.3, and $\Omega_{\wedge}=$0.7.

\section{Data Sets \& Sample Definition}
\label{sec:subsamps_sec}
\subsection{Initial Selection}
To construct our sample of galaxies, we make use of the full xCOLD GASS \citep{saintonge17}, xGASS \citep{catinella18} and ALFALFA $\alpha$.100 \citep{haynes18} data sets, providing IRAM 30m CO(1-0) observations for 532 galaxies and Arecibo observations for a combined 32,111 galaxies, respectively. Observations of CO(1-0) at 3\,mm allow us to trace directly the cold (T$\lesssim$100 K) molecular gas used for star formation in galaxies in the local Universe, whilst the \HI\ 21cm line in emission traces more diffuse, cool (T$\lesssim$5,000 K) Hydrogen gas. We begin by defining sub-samples of galaxies which are most likely to host outflows, based on their position on the SFR-M$_{*}$ plane and inclination. All xCOLD GASS and xGASS galaxies have SDSS optical counterparts and therefore catalogued stellar masses and SFRs, and 18,727/31,502 galaxies from ALFALFA also have optical counterparts matching the coordinates of spectroscopic galaxies in the MPA-JHU DR7 catalog\footnote{https://wwwmpa.mpa-garching.mpg.de/SDSS/DR7/}, within 10$''$ of the target. Here we use the SFRs and stellar masses as measured in the MPA-JHU catalog. For both our CO and \HI\ data, we select galaxies lying above the lower limit of the MS, as defined by \citet{saintonge16}, with M$_{*}\geqslant$10$^{10}$ M$_{\odot}$, and inclinations less than or equal to 60$^{\circ}$. Additionally, we select only star-forming galaxies based on a \citet{ka03} BPT cut (therefore rejecting AGN-dominated objects). Although recent results from \citet{rb19} suggest a slightly lower inclination cut (i.e., $i<$50$^{\circ}$), outflows have been observed with inclinations up to $i\sim$60$^{\circ}$ and the modest sample size of xCOLD GASS motivates us to maximise our outflow sample where possible. The selection criteria we impose results in a sample of 69 star-forming galaxies from xCOLD GASS and 1,487 \HI\ galaxies from ALFALFA and xGASS, and we refer to these as our parent samples. Additionally, we also make use of 78 high mass, star-forming and low-inclination ($i<$50$^{\circ}$) MaNGA DR15 IFU galaxies selected to show \nad\ outflows by \citet{rb20}, with which we can constrain the neutral and ionised contributions to outflows from \nad\ absorption and ionised emission. The motivation for choosing MaNGA rather than SDSS optical observations is due to the fact that IFU observations allow us to probe higher S/N and larger areas for each galaxy, which have shown to contain non-negligible contributions of outflowing gas that would be missed by a 3$''$ fiber \citep{rb20}.

\subsection{Removal of Confused Sources}
\label{subsec:confrem}
Spectroscopic confusion in radio observations from single dish facilities is an important concern, and in particular for stacking and outflow studies, since real signal from nearby galaxies at similar velocities can mimic the signatures of outflowing gas. Although the IRAM 30m telescope has a FWHM$\sim$22$''$ at 3\,mm, the observations of \HI\ used in this study are observed with the 3.8$'$ Arecibo beam, significantly enhancing the potential for confusion. Thus, we remove all galaxies from the xGASS catalogs which are flagged as confused and further remove objects where such information is not available using the following method: 

\begin{enumerate}
    \item We use galaxy coordinates from the MPA-JHU catalog to identify galaxies within a separation of $\leqslant$3$\times$FWHM from a given target galaxy (assuming a beam of FWHM=22$''$ for observations with the IRAM 30m and FWHM=3.8$'$ for observations with the Arecibo telescope) but farther than 10$''$ away from the target galaxy to avoid including it as a contaminant.
    \\
    \item For each galaxy where a nearby optical galaxy is found, the velocities of the galaxies are compared. Because our stacking process (described below) requires us to normalise the spectrum by the width of the emission line, applying a fixed velocity cut could miss a significant portion of confused galaxies. As such, we require galaxies to be within 5$\times$ the width of the target's emission line for them to be contaminating.
    \\
    \item Due to our large separation requirements, many of the flagged target galaxies have ``contaminating'' galaxies outside of the FWHM of the telescope beam (where the sensitivity drops from 50\% to effectively 0\%), and as such are likely to provide negligible contribution to the CO or \HI\ emission. Thus, we compare derived gas masses of the contaminants to the catalogued values of the target galaxy via scaling relations of molecular and atomic gas fractions and specific SFR \citep{saintonge17,catinella18}. If a neighbouring galaxy contributes $\geqslant$10\% of the target's measured CO or \HI\ gas mass, the target galaxy is deemed to be subject to confusion and discarded.
\end{enumerate}

This procedure is performed over all xCOLD GASS, xGASS and ALFALFA galaxies, but also between the samples themselves so that even galaxies without optical counterparts are accounted for. In total, we flag 3/69 CO and 303/1,487 \HI\ galaxies as confused and discard them from any remaining analysis. Finally, we visually inspect the spectra of the remaining galaxies for evidence of baseline, artefact or radio frequency interference (RFI) issues and remove 1 CO and 107 \HI\ sources with suspect spectra. In total, this results in 78 optical galaxies, 65 CO galaxies and 1,077 \HI\ galaxies which form the final samples for this analysis. 37 of the galaxies from the final CO and \HI\ samples overlap within separations of $<$2$''$ and $|\Delta z|<$5$\times$10$^{-4}$, whilst 2 and 8 galaxies from the respective samples also overlap with the MaNGA sample. None of the targets have overlapping observations for all 3 tracers using the surveys in this study. We show the CO(1-0) and \HI\ spectra, as well as the SDSS postage stamp image, of 10 randomly chosen galaxies in our final samples with overlapping observations in Figure \ref{fig:xCGpostage}, and present the mean galaxy properties of each sample in Table \ref{tab:meanprops}.

\begin{table*}
    \centering
    \begin{tabular}{lccccc}
      \hline
        Sample & log M$_{*}$ & log SFR & Redshift & Inclination & Nr. of Galaxies \\
          & [M$_{\odot}$yr$^{-1}$] & [M$_{\odot}$] &  & [degrees] &  \\
        \hline
        MaNGA DR15 & 10.75$\pm$0.28 & 0.80$\pm$0.40 & 0.046$\pm$0.024 & 35$\pm$10 & 78\\
        xCOLD GASS & 10.48$\pm$0.33 & 0.57$\pm$0.35 & 0.036$\pm$0.008 & 44$\pm$11 & 65\\
        xGASS+ALFALFA (star-forming) & 10.41$\pm$0.32 & 0.40$\pm$0.50 & 0.038$\pm$0.011 & 41$\pm$13 & 1,077\\
        xGASS+ALFALFA (passive) & 10.70$\pm$0.34 & -0.99$\pm$0.28 & 0.029$\pm$0.012 & 45$\pm$20 & 341\\
        \hline
    \end{tabular}
    \caption{The mean integrated galaxy properties of each sample defined in Section \ref{sec:subsamps_sec} of this work and shown in Figure \ref{fig:subsamps} (dark blue and red points). The errors quoted here are the 1$\sigma$ standard deviation over each sample.}
    \label{tab:meanprops}
\end{table*}

\begin{figure*}
\center
 \includegraphics[width=0.99\textwidth]{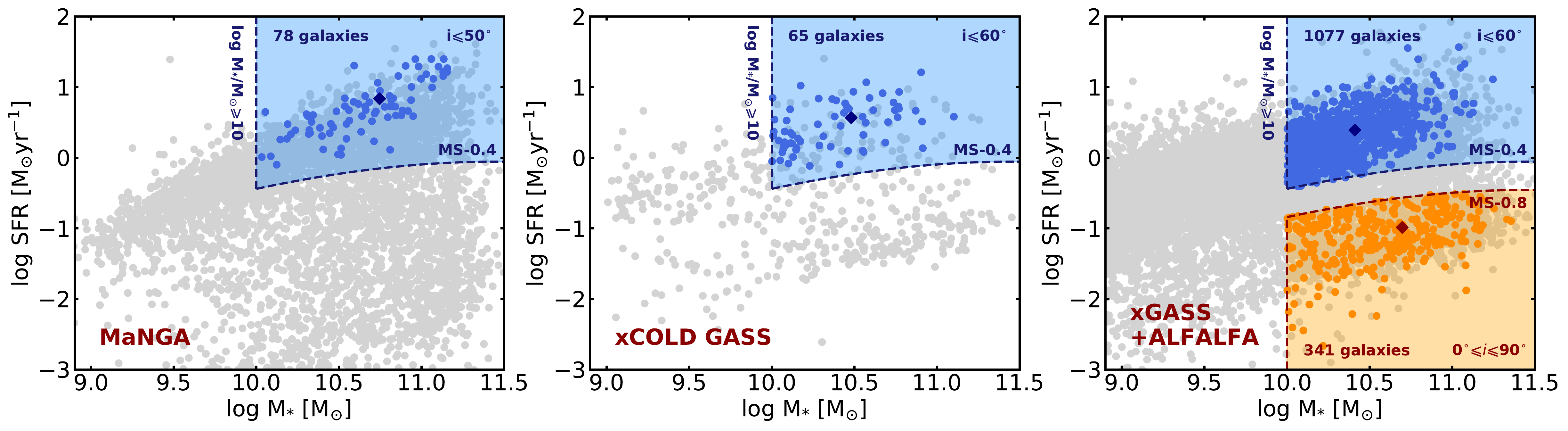}
  \caption{The distribution of galaxies from the full MaNGA DR15 (left), xCOLD GASS (middle) and combined xGASS+ALFALFA $\alpha$.100 (right) data releases (gray points) and our selected star-forming sample of high mass galaxies with low inclinations (blue points) across the SFR-M$_{*}$ plane. The blue regions and dashed lines mark our selection limits across the plane for star-forming galaxies likely to host outflows, with the SFR limit set by the definition of the MS by \citet{saintonge16}. A navy diamond in each plot marks the mean SFR-M$_{*}$ position of each selected star-forming sample. The orange region and dark red lines in the right panel represent the selection limits for our sample of \HI\ passive across all inclinations and galaxy types, where outflows are not expected to be seen.}
 \label{fig:subsamps}
\end{figure*}

\begin{figure*}
\center
 \includegraphics[width=0.99\textwidth]{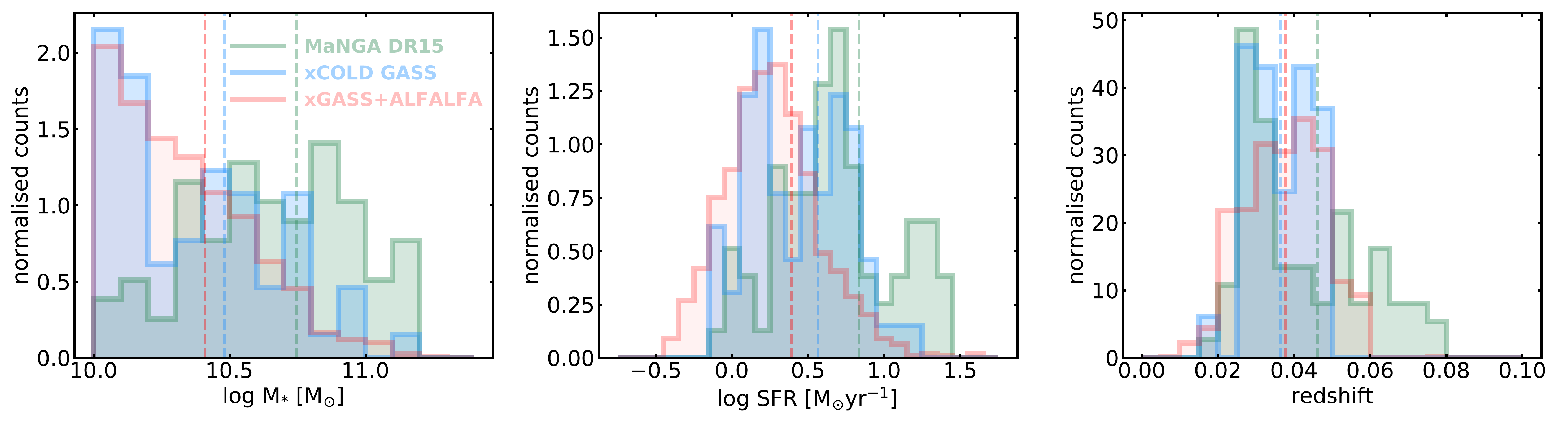}
 \caption{The normalised stellar mass (left), SFR (middle) and redshift (right) distributions of our three final samples derived from the MaNGA DR15 (green), xCOLD GASS (blue) and xGASS+ALFALFA $\alpha$.100 (red) surveys. The dashed lines represent the mean values for each distribution.}
 \label{fig:distributions}
\end{figure*}

The positions of our final selected samples of optical, CO and \HI\ galaxies on the SFR-M$_{*}$ plane are shown in Figure \ref{fig:subsamps} and their stellar mass, SFR and redshift distributions are show in Figure \ref{fig:distributions}, where one can see that the maximum difference between the median SFRs and stellar masses of each star-forming sample is 0.4 dex and 0.36 dex, respectively.

\section{Stacking Methods \& Validation}
\subsection{Stacking Procedure}
In order to obtain high S/N spectra with which to search for broad component emission imprinted by outflows, we opt for a velocity-normalised stacking approach, where the emission from each galaxy is first normalised by its systemic width prior to being added to a stack. Since the conventionally defined W50 (or FWHM) parameter may not always encapsulate all of the systemic flux of the line and its use could lead to false-positive outflow detections from leaked emission, we opt to guard against this by measuring a full width (FW) of the line and remeasure its velocity, thereby encapsulating virtually all of the systemic emission. For spectra where an emission line is not detected, we use the optical SDSS redshift and the Tully-Fisher relation from \citet{tiley16} to estimate a velocity and width, respectively. We explore the suitability of this approach in Section \ref{sec:reliab}. Finally, because \HI\ spectra can often display particularly noisy or asymmetrical spectra where the width of the line is not well constrained, we define a sample of 220 “pristine” \HI\ galaxies for which the emission is of significant S/N with well defined widths.

After deriving FWs and remeasured velocities for both the CO and \HI\ spectra, each emission line is shifted to its rest frame before being normalised by its FW and interpolated over a common array. Because our goal is to derive mass outflow rates of potential outflowing gas, we stack each spectrum in ``gas mass density units'', that is to say:

\begin{itemize}
	\item For CO(1-0): each spectrum is multiplied by all the standard factors in the \citet{solomon97} prescription (without integrating over the normalised velocity space, v$_{\text{norm}}$) and an assumed conversion factor for normal star-forming galaxies, $\alpha_{\text{CO}}$=4.35 M$_{\odot}$ (K\,km\,s$^{-1}$\,pc$^{-2}$)$^{-1}$, which includes a correction for the abundance of Helium (\citealt{accurso17}, and references therein).

	\item For \HI\ : each spectrum is converted to an \HI\ gas mass density via:
	\begin{equation}
    \label{eq:mhi}
    M_{\text{\HI\ }}/\text{v}_{\text{norm}} = 2.356\times10^{5} \bigg(\frac{D}{\text{Mpc}}\bigg)^{2} \bigg(\frac{F_{\text{HI}}}{\text{Jy}}\bigg)
    \end{equation}
\end{itemize}

The resulting spectra therefore have units of M$_{\odot}$/v$_{\text{norm}}$. Each spectrum is subsequently added to the stack and the final stacked spectrum is taken as the mean over all galaxies in the stack. The errors associated with the final spectrum in all of our stacks are taken as the combination of bootstrapped sampling errors (with replacement) and the average flux uncertainties of each spectrum. \\

For each of the 78 MaNGA galaxies, we first stack spaxels within a 0.5\,R$_{e}$ galactocentric radius, since this is shown to be a rough limit where \nad\ is found in absorption \citep{rb20}. Each spectrum from a given spaxel is put through the steps outlined in \citet{rb19} and we refer the reader to that paper for details, although provide a brief summary here: the spectrum is first corrected for galactic extinction using the \citet{schlegel98} dust maps and a \citet{odonnell} Milky Way extinction curve, prior to being shifted to the rest-frame (using an H$\alpha$ redshift; \citealt{talbot18}) and interpolated over a common wavelength array. The resulting spectra are then converted to a luminosity and summed together to obtain a single spectrum akin to a single deep exposure over the entire 0.5\,R$_{e}$ of each galaxy. To obtain the mean 0.5\,R$_{e}$ galaxy spectrum, we then proceed to stack the summed spectra over all galaxies, with two slightly different approaches, depending on the tracer of interest. For \nad\, each spectrum is normalised by its mean flux between 5450 \AA\ and 5550 \AA\ (where the spectrum is free from absorption or emission lines) and added to the stack, which is then averaged and subsequently fit with a continuum model from the widely-used pPXF code \citep{cappellari17} with which we divide the average galaxy spectrum to obtain the \nad\ ISM contribution. According to the method described by \citet{rb19}, the \nad\ profile is then fit with a single fixed and a single offset profile to determine whether an outflow is present, and subsequently a three-component model (consisting of blueshifted absorption, systemic absorption, and redshifted emission components) if an outflow is detected. For the ionised gas, on the other hand, we fit and subtract a pPXF continuum model from each galaxy spectrum prior to adding it to the stack, then take the average of the stacked residual spectra. This is to ensure the final spectra have the correct units and the weighting of the optical spectra are the same. In this study we use H$\alpha$ as the main tracer for the ionised gas. The primary reasons for this are (i) due to the brightness of the emission line and accompanying [\NII] doublet with which to constrain fits to the data, (ii) the ability to avoid making additional (and unconstrained) metallicity assumptions as with e.g., [\OIII], and (iii) in order to compare to the studies of \citet{fluetsch19} and \citet{gallagher19} who also use H$\alpha$ as the primary ionised gas tracer for local star-forming galaxies. The final residual luminosity of the spectrum is thus converted to ``outflow gas mass units'' via the relation,

\begin{equation}
\label{eq:ionmass}
M_{\text{H}\alpha} = \frac{1.4\,m_{\text{H}}\,L_{\text{H}\alpha,\text{out}}}{\gamma_{\text{H}\alpha}\,n_{e}},
\end{equation}

where $\gamma_{\text{H}\alpha}$ is the H$\alpha$ emissivity at a temperature $T$=10$^{4}$ K ($\gamma_{\text{H}\alpha}$=3.56$\times$10$^{-25}$ erg cm$^{3}$ s$^{-1}$) and $n_{e}$ is the local electron density of the outflow. Such an approach allows us to simulate a single long exposure for each MaNGA galaxy (similar to our (sub)mm and radio observations), maintain high S/N and maximise the area of the galaxy probed.

\subsection{Reliability of Stacking Methods and Validity of Broad CO and HI Emission}
\label{sec:reliab}
Guarding against false-positive outflow detections is paramount to gaining an accurate understanding of their multiphase nature. As such, a number of factors must be taken into account, which could all mimic a broad component.

A first consideration is that broad emission is the result of nearby sources that confuse the spectra of our targets. However, the rate of spectroscopic confusion is generally small (less than $\sim$15\% for the ALFALFA $\alpha$.40 data release; \citealt{jones15}) and we already account for this possibility by removing galaxies subject to confusion in Section \ref{subsec:confrem}.

A second consideration is to ensure any resulting broad CO or \HI\ emission would not be an artefact of our stacking approach. To test this, we create a stack of CO and \HI\ spectra where, for each spectrum going into the stack, we place the systemic component of the emission line - as defined by our measured FW - in a random region (within 1,000 \kms of the systemic velocity) of the spectrum and replace the previous location of the emission line with the (noise) portion of the spectrum now taken up by the emission line, thereby redefining the central velocity of the line. It follows that, since only systemic emission is selected, a broad component would only appear as a result of our stacking process. After taking the mean of the stack, we inspect whether a broad component arises from our procedure: neither stack (CO or \HI) reveals any evidence of a broad component and therefore validates our stacking method.

\begin{figure}
\center
 \includegraphics[width=1.\columnwidth]{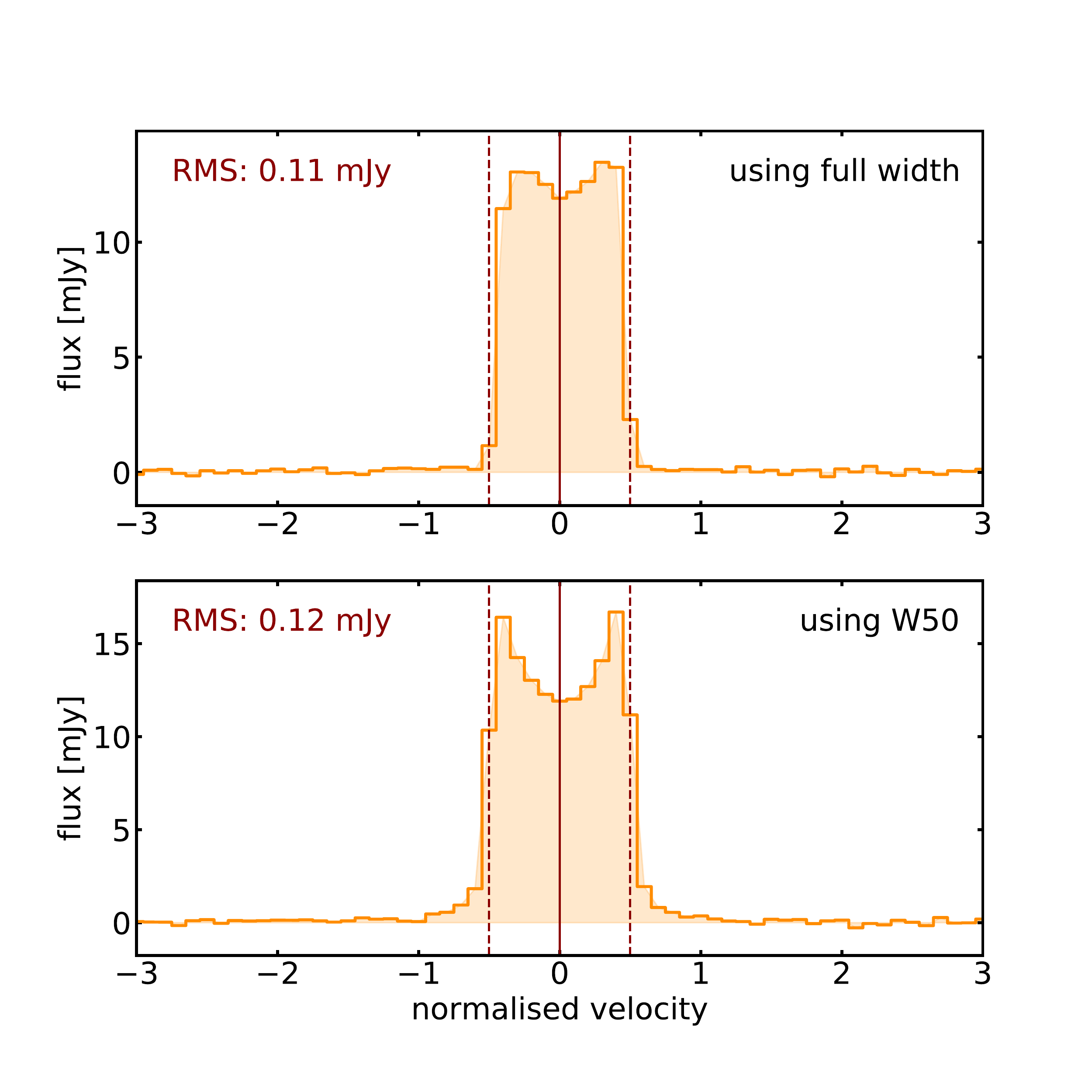}
 \caption{\HI\ stacks of high mass, passive galaxies (including AGN) over the full range of inclinations, where outflows are not expected to be seen. Stacks are created via a Monte Carlo sampling of 80\% of the sample, repeated 300 times. The top panel shows the stacked spectrum using our measured linewidths and velocities and the bottom panel repeats the same process using the catalogued W50 values. A clear difference is seen between the two, with the former stack displaying no signatures of broad emission, suggesting all of the systemic emission is encapsulated within our measurement, while the latter displays significant broadening at high normalised velocities, indicative of leaking systemic emission. Such a comparison is imperative in a search for outflows, since the broad emission could be interpreted as outflowing signal in star-forming samples.}
 \label{fig:passivestacks}
\end{figure}

A third - and most important - consideration is to ensure that the systemic flux of each galaxy is completely encapsulated by the width of the line, in order that systemic emission does not ``leak'' into broad emission once normalised by the width of the line. To test this, we define a sample of \HI\ -detected, high mass, passive galaxies of all inclinations and galaxy types from xGASS and ALFALFA, where we do not expect to observe outflows and therefore any potential broad emission would be the result of systemic emission not captured by our measured width. Specifically, this sample of galaxies is defined as having log\,M$_{*}$/M$_{\odot}\geqslant$10 and log\,SFR/M$_{\odot}$yr$^{-1}\leqslant$MS-0.8\,dex and also includes AGN. Just as for galaxies in our star-forming samples, for each galaxy we redefine the central \HI\ line velocity and measure the FW and remove any object subject to confusion and/or contaminated by artefacts or RFI. This results in a sample of 341 passive galaxies with useable spectra, which are shown in Figure \ref{fig:subsamps}. We subsequently adopt a Monte Carlo approach to randomly select and stack 80\% of this sample 300 times, where the final mean spectrum is taken as the average over all Monte Carlo iterations. Additionally, to validate our choice of using the FW and remeasured velocity the CO and \HI\ spectra used in this study, we perform the same stack using the velocities and FWHM (W50) derived from the xGASS and ALFALFA catalogs. Both stacks are presented in Figure \ref{fig:passivestacks} and reach depths of RMS$\sim$0.11-0.12 mJy, measured between normalised velocities of 1.5$\leqslant|$v$_{\text{norm}}|\leqslant$3 either side of the emission line and normalised velocity channels of 0.1.

\begin{figure}
\center
 \includegraphics[width=\columnwidth]{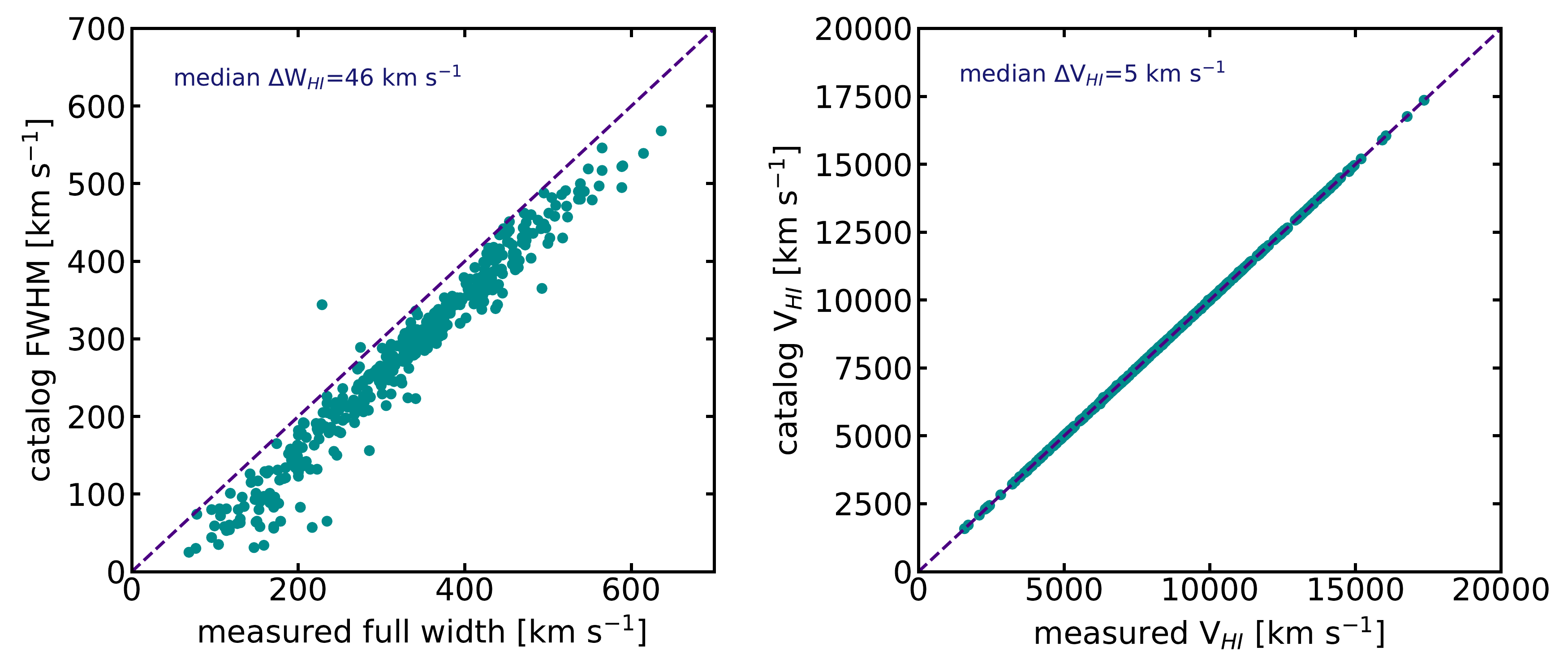}
 \caption{A comparison of the catalogued \HI\ FWHMs (W50) versus our measured full linewidths (left) and the same for \HI\ velocities (right). Our linewidths are systematically larger than the catalogued values, with a median difference of $\sim$46 \kms. The difference between \HI\ velocities is much smaller, with a median difference of only $\sim$5 \kms. In both plots, the dashed purple line denotes a 1:1 relation. The difference in \HI\ linewidths is crucial in our search for broad emission outflows, since an underestimation of the linewidth can result in leaked systemic emission which can mimic the signature of outflows.}
 \label{fig:comps}
\end{figure}

From the top panel of Figure \ref{fig:passivestacks}, we immediately note a complete lack of broad emission, suggesting our measured widths encapsulate all (and serve as reliable measurements of) the systemic emission of the line. However, from the bottom panel of the same figure, we note that significant broad wings appear in the high S/N spectrum, indicating that the catalogued W50 values are inadequate tracers of systemic emission for our purposes. Thus, using the W50 widths for our star-forming sample could result in broad emission that could falsely be interpreted as an outflow component and highlights the precarious nature of searching for broad components in emission spectra. We illustrate the differences between our measured widths and velocities compared to the catalogued values in Figure \ref{fig:comps}, with median differences of 46 \kms and 5 \kms, respectively. The tests performed here validate our choice of remeasuring the line widths and velocities, for which we find the catalogued W50 values underestimate the systemic component by (a median) $\sim$46 \kms, and ensure potential outflowing signal is not due to artefacts of our stacking procedure or measurements.

\section{Results}
\label{sec:res}
\subsection{Neutral and Ionised Gas Outflows}
\begin{figure*}
\center
 \includegraphics[width=\textwidth]{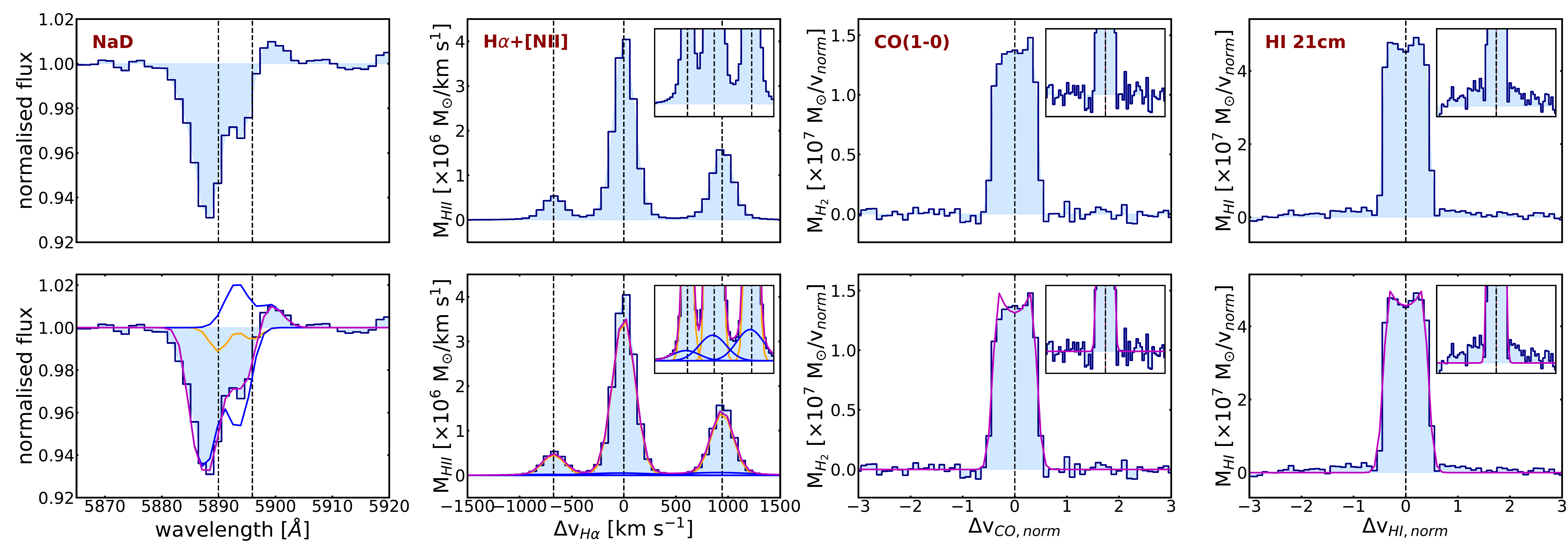}
 \caption{The final stacked spectra of our \nad\ (left column), H$\alpha$+[\NII] (middle left column), CO(1-0) (middle right column) and \HI\ 21cm (right column) tracers from the MaNGA DR15, xCOLD GASS and xGASS+ALFAFA $\alpha$.100 surveys. The top row shows the stacked spectra only and the bottom row shows the stacked spectra with their best fit models. Stacked flux is outlined by the navy line and light blue fill, orange lines mark fits of a systemic component, blue lines mark outflowing components, and purple lines mark the total fit. The inset plots in the emission line columns are zoomed in portions of the respective spectra to better display the possible presence and fits of broad outflow components at lower flux intensity.}
 \label{fig:finalspecs}
\end{figure*}

In Figure \ref{fig:finalspecs} we present the results of our stacking approach for \nad, H$\alpha$+[\NII], CO(1-0) and \HI\ 21cm gas tracers from our MaNGA, xCOLD GASS and xGASS+ALFALFA samples. We begin by examining the results seen in the \nad\ profile, which displays significant blueshifted absorption and redshifted emission characteristic of a P-Cygni profile, which is an unambiguous signature of outflows. A three-component fit to the line (systemic, blueshifted absorption, and redshifted emission; see \citealt{rb19} for details) reveals a blueshifted outflow component with velocity $|$v$_{\text{out}}|$=131 \kms (160 \kms\ after correcting for a mean inclination of $i\sim$35$^{\circ}$) and Hydrogen column density of $N(H)$=10$^{21.33}$ cm$^{-2}$ (using Equation 7 and assumptions in \citealt{rb19}). For a spherically-symmetric, mass conserving outflow subtending a solid angle less than 4$\pi$ and radius of 5\,kpc, the mass outflow rate can be expressed as:

\begin{equation}
\label{eq:finaldM}
\begin{split}
\dot{M}_{\text{out}} = 115\sum\,\bigg(\frac{C_{\Omega}}{0.4}\,C_{f}\bigg)\,&\bigg(\frac{r}{10\,\text{kpc}}\bigg) \\
&\times\bigg(\frac{N(\text{H})}{10^{21}\,\text{cm}^{-2}}\bigg)\bigg(\frac{|\Delta\,v|}{200\,\text{km\,s}^{-1}}\bigg) \,\text{M}_{\odot}\,\text{yr}^{-1},
\end{split}
\end{equation}

Using the above equation and derived \nad\ parameters, we find an inclination-corrected neutral mass outflow rate of $\dot{M}_{\text{NaD}}$=7.55$\pm$7.20 M$_{\odot}$\,yr$^{-1}$, consistent with other neutral gas outflow rates derived for normal galaxies at $z\sim0$ \citep{martin12,rubin14}. The uncertainties quoted here include those from the fit parameters and those of the mean inclination. The redshifted emission is much less stark than the blueshifted absorption, although is a clear indication of resonant re-emission coming from the backside of the galaxy disk (i.e., a receding outflow) and provides compelling evidence for the presence of outflows in our selected galaxies.

We next inspect the profiles of the H$\alpha$ and [\NII] emission lines. The H$\alpha$ and [\NII] lines are close enough to each other ($<$1000 \kms) to cause some blending of the profiles, however it is still possible to derive accurate fits of the lines due to the high S/N of the spectra. Thus, we aim to fit all three of the aforementioned lines simultaneously, in order to extract as much information as possible from direct probes of the state of the gas, as well as to limit degenerate fits of broad components \citep{newman12a,fs14,davies19,fs19}. To determine whether outflowing gas is seen in ionised gas tracers, we fit the combined H$\alpha$+[\NII] emission with both a single Gaussian fit for each line and a double-Gaussian fit consisting of a systemic (narrow) and a broad Gaussian profile for each line. For the systemic fits, we allow the amplitude of each line to vary but fix the amplitudes of the [\NII] lines to their intrinsic ratio of [\NII]$\lambda$6548/[\NII]$\lambda$6583 = 0.326 \citep{morton91}, and assume the same width for all lines which is allowed to vary from 0$\leqslant$FWHM$\leqslant$800 \kms. For the broad components, we make the same assumptions as the systemic components, but allow the amplitudes to reach no higher than the systemic counterpart and require the width to be larger than the systemic components. Additionally, we also allow a velocity offset of the broad components (the same for each component) of $\Delta$v$\pm$500 \kms. A comparison of the two fits (single versus double component) with a Bayesian Information Criterion (BIC) reveals that the extra four free parameters in the double component fit are justified, and as such an ionised outflow is detected. We measure an outflow velocity (v$_{\text{out}}$=FWHM/2 + $|\Delta$v$|$) of 439 \kms\ (534 \kms\ after inclination correction), more than a factor of 3$\times$ larger than the velocity found by the blueshifted absorption in \nad. For an estimation of the electron density within the outflow, we adopt a low value of $n_{e}$=50 cm$^{-3}$ \citep{newman12a,fs19}, although we note this is likely to maximise the resulting gas mass. We subsequently derive a mass outflow rate from fits to the emission lines as follows:

\begin{equation}
\label{eq:emdmdt}
\dot{M}_{\text{out}} = \frac{\text{v}_{\text{out}} M_{\text{out}}}{R_{\text{out}}},
\end{equation}

where v$_{\text{out}}$ is the inclination-corrected outflow velocity, $M_{\text{out}}$ the instantaneous gas mass of the outflow, and $R_{\text{out}}$ is the assumed radius of the outflow (5 kpc). Thus, integrating over the broad component of H$\alpha$ and using Equation \ref{eq:emdmdt}, we infer a mass outflow rate of 0.10$\pm$0.02 M$_{\odot}$\,yr$^{-1}$.

\subsection{Molecular and Atomic Gas Outflows}
Finally, we turn our attention to the stacked CO(1-0) and \HI\ 21cm emission, which serve as tracers of fuel for star formation. At first glance, we observe no significant evidence for broad wings either side of the double horned (DH) profiles. To confirm this, we fit both the CO and \HI\ profiles with a single DH profile, composed of a parabolic function accompanied by two equidistant (and identical but mirrored) half-Gaussians used to describe the low- and high-velocity edges of the DH (the ``Gaussian Double Peak'' function, described in detail in \citealt{tiley16}), and a two-component profile consisting of a DH profile and broad Gaussian. The number of free parameters for these fits are 4 and 7, respectively. For the DH profile, we allow the height of the parabola and peak of the two Gaussians to reach the maximum observed flux (within errors), whilst restricting the half width of the parabola to 0$\leqslant \Delta$v$_{\text{norm}}\leqslant$1.5 and the width of the Gaussians to 0$\leqslant\sigma_{\text{norm}}\leqslant$2. For the broad Gaussian, we allow the amplitude to reach no higher than the central parabola of the DH component and restrict the width to greater than that of the DH profile but less than $\sigma_{\text{norm}}\leqslant$5.0. Furthermore, we don't allow for a velocity offset.

Our fits to the CO(1-0) emission and a comparison of the BIC values determine that a two-component fit is unjustified, suggesting that no outflow is present, despite our stack reaching a (RMS) depth of log M$_{H_{2}}/\text{v}_{\text{norm}}\sim$5.6 M$_{\odot}$/v$_{\text{norm}}$. The \HI\ emission, on the other hand, displays tentative broad emission on either side of the profile, with a measured RMS of log M$_{\text{HI}}/\text{v}_{\text{norm}}\sim$5.8 M$_{\odot}$/v$_{\text{norm}}$. Both RMS values are measured in the same way as described in Section \ref{sec:reliab}. A two-component fit to the profile quantitatively confirms the broad emission via a BIC ratio (BIC$_{\text{1-comp}}$/BIC$_{\text{2-comp}}$), however the amplitude of the broad Gaussian is detected only to $<$2$\sigma$ when compared to the noise of the spectrum. If the stacking process is repeated over our 220 pristine galaxies, where the width and velocity of the line are better constrained, we reach a sensitivity of log M$_{\text{HI}}/\text{v}_{\text{norm}}\sim$6.0 M$_{\odot}$/v$_{\text{norm}}$ but find the tentative broad emission disappears. Despite having performed extensive tests to significantly limit the potential contribution from confusion, leaked systemic emission, or spectral artefacts, we cannot completely rule out some contamination in our main stack and the emission is sufficiently faint that we cannot make robust claims as to its source. Additional, dedicated observations would be required to confirm such emission as outflowing gas. Finally, in Figure \ref{fig:subsamps} we note that the mean stellar mass of the combined \HI\ sample is somewhat lower than the MaNGA sample, where we observe outflows. Since this has been shown by several studies to be a considerable factor in optical outflow detections \citep{chen10,concas19,rb19}, as an additional test we perform the same stack over higher mass (log M$_{*}\geqslant$10.5 M$_{\odot}$) galaxies only. 263 galaxies result from this cut, and the absolute differences in the mean properties of the MaNGA and higher mass \HI\ sample are far better matched, with differences of $\Delta$log M$_{*}$=0.027 M$_{\odot}$, $\Delta$log SFR=0.168 M$_{\odot}$yr$^{-1}$ and $\Delta z$=0.006. In the resulting stack we find very tentative wings of \HI\ emission, which could be suggestive of outflowing gas and a comparison of BIC values from a single DH fit or DH+Gaussian fit reveals the latter is preferred and a supposed detection is found. However, such ``wings'' are extremely faint and in subtracting the single DH fit from the spectrum and comparing the residual wings to the noise in the spectrum, we find they are only detected to 1.4$\sigma$. As such we cannot make robust claims as to an \HI\ outflow detection.

\begin{figure}
\centering
 \includegraphics[width=\columnwidth]{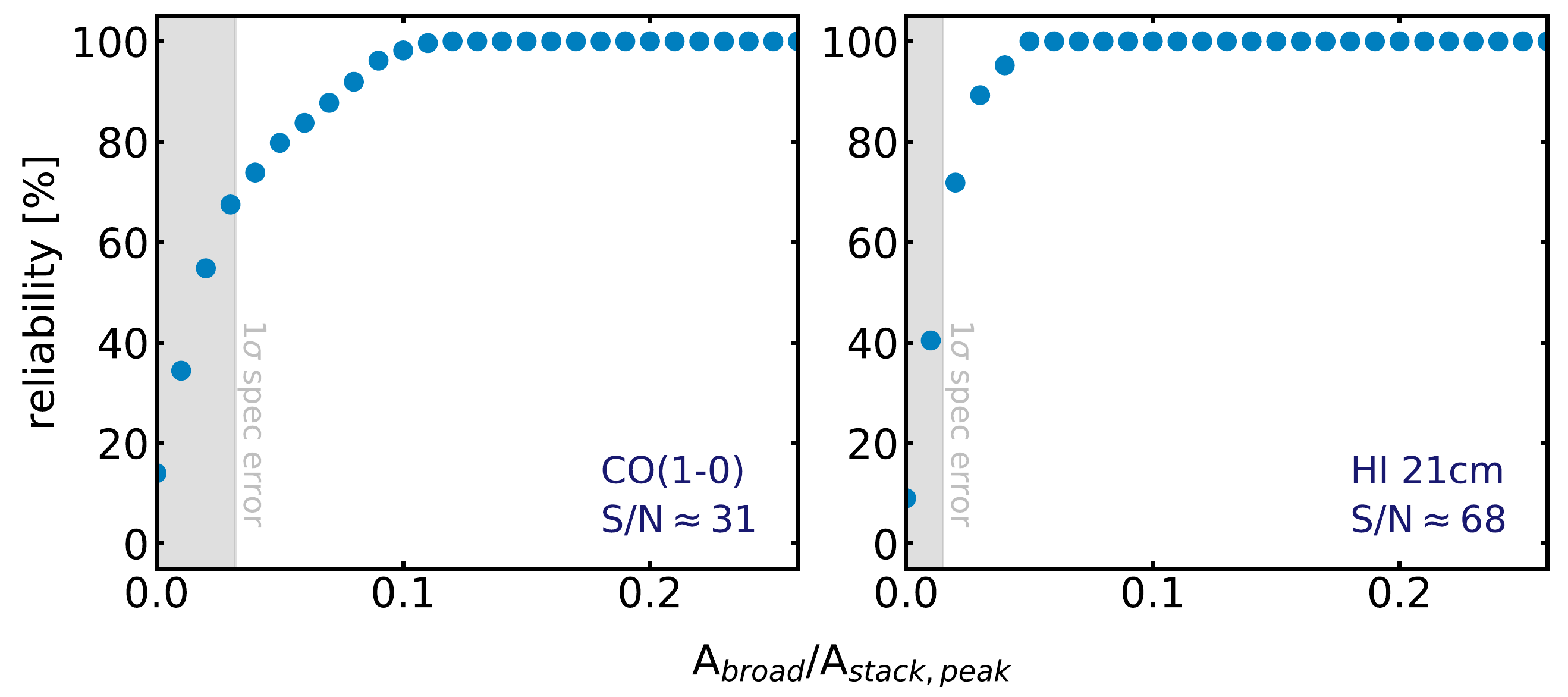}
 \caption{The reliability of our code's outflow detections at various amplitude ratios for simulated spectra of the same S/N of our stacked CO (left) and \HI\ (right) spectra, given an assumed outflow velocity of 200 \kms. The 1$\sigma$ error of each spectrum relative to the amplitude of the stack is highlighted in gray. The lowest outflow amplitude ratio (compared to the peak flux of the total spectrum) we can measure whilst maintaining 90\% reliability is A$_{\text{broad}}$/A$_{\text{stack,peak}}$=0.075 (CO) and A$_{\text{broad}}$/A$_{\text{stack,peak}}$=0.031 (\HI).}
 \label{fig:reliability}
\end{figure}

Given the non-detections described above, a certain degree of inference is necessary to place upper limits on the mass outflow rate of molecular and atomic gas with our high S/N spectra. To do this, we run simulations of completeness and reliability for our code to pick up faint outflow signatures of an assumed velocity, using synthetic spectra containing both a DH profile and a broad Gaussian combined with a level of Gaussian noise matching that measured in the stacked CO and \HI\ spectra. Our procedure is as follows: first, in order to derive an independent measurement of systemic emission where we are confident outflow signal is absent, we fit our passive galaxy stack from Section \ref{sec:reliab} with a single DH profile as described above and use this as our profile for systemic emission. Next, we construct a broad Gaussian component by assuming an outflow velocity of 200 \kms\ (the choice is based on the low outflow velocities found in the literature for normal and starburst galaxies at $z\sim$0, e.g., \citealt{chen10,sugahara17,concas19,fluetsch19,rb19}) which we compare to the median FW of the spectra going into the CO and \HI\ stacks (FW$\approx$284 \kms\ and FW$\approx$275 \kms\ for CO and \HI, respectively) in order to determine the corresponding normalised outflow velocity. The normalised FWHM of the Gaussian is given by FWHM$_{\text{broad,norm}}$ = v$_{\text{out,norm}} \times$2 and we determine the normalised outflow velocity from v$_{\text{out,norm}}$ = v$_{\text{out}}$/HW$_{\text{median,CO/HI}}$, where HW$_{\text{median,CO/HI}}$ is the median half-width going into the stacks. The broad Gaussian is then constructed assuming a variety of amplitude ratios (A$_{\text{broad}}$/DH$_{\text{sys,peak}}$, where A$_{\text{broad}}$ is the amplitude of the broad component and DH$_{\text{sys,peak}}$ is the peak flux of the systemic DH profile) and added to the systemic DH profile to create the total spectrum. Random Gaussian noise is then added to the spectrum to match the measured value of the CO and \HI\ spectra and the final profile is subsequently fit with a single DH profile and combined DH+Gaussian profile, as described above but allowing full range to the width of the fit Gaussian profile. The process of Gaussian noise addition and subsequent fitting is repeated 100 times. We subsequently assess the completeness and reliability of our fits, to determine the minimum amplitude of the outflow component that our code can reliably measure. We find in all cases the DH+Gaussian profile is preferred, suggesting 100\% completeness, however many of these (particularly at low Gaussian amplitudes) have measured Gaussian FWHM smaller than the DH profile FWHM (i.e., the Gaussian is fitting noise in the centre of the spectrum, not outflow signal). We consider fits with FWHM$_{\text{broad}}$/FWHM$_{\text{DH}}<$1 as false-positive detections and the reliability is therefore defined as the fraction of fits without false-positives. A plot of reliability for both the simulated CO and \HI\ spectra is shown in Figure \ref{fig:reliability}.

We find the minimum outflow amplitude (compared to the peak flux of the stacked spectrum, A$_{\text{stack,peak}}$) we can measure with $>$90\% reliability is A$_{\text{broad}}$/A$_{\text{stack,peak}}$=0.075 for CO and A$_{\text{broad}}$/A$_{\text{stack,peak}}$=0.031 for \HI, corresponding to outflow gas masses of log M$_{\text{gas}}$/M$_{\odot}\sim$8.68 and log M$_{\text{gas}}$/M$_{\odot}\sim$8.82, respectively. Using Equation \ref{eq:emdmdt}, we derive upper limits on the mass outflow rate of molecular and atomic gas of $\dot{M}_{\text{H}_{2}}<$19.43\,M$_{\odot}$yr$^{-1}$ and $\dot{M}_{\text{HI}}<$26.72\,M$_{\odot}$yr$^{-1}$, respectively, and summarise the parameters of the above findings in Table \ref{tab:multip}.

\begin{table}
	\centering
	\begin{tabular}{lcccc}
      \hline
        Quantity & CO(1-0) & \HI\ 21cm & \nad\ & H$\alpha$ \\
        \hline
        v$_{\text{out}}$ [\kms] & -- & -- & 131$\pm$8 & 439$\pm$11 \\
        v$_{\text{out}}$/cos($i$) [\kms] & 200$^{a}$ & 200$^{a}$ & 160$\pm$10 & 534$\pm$14 \\
        $\dot{M}_{\text{gas}}$ & $<$19.43 & $<$26.72 & 7.55$\pm$7.20 & 0.10$\pm$0.02 \\
        $\eta$ [$\dot{M}$/SFR] & $<$5.27 & $<$10.86 & 1.10$^{+1.46}_{-1.10}$ & 0.02$\pm$0.01 \\
        \hline
	\end{tabular}
	\caption{The mean properties and parameters of our stacked spectra used to derive multiphase mass outflow rates in this work. \newline
	$^{a}$Assumed outflow velocities.}
	\label{tab:multip}
\end{table}

\section{Discussion}
\subsection{Towards a Total, Multiphase Mass Outflow Rate}
\label{subsec:multiphasedmdt}
\begin{figure*}
\center
 \includegraphics[width=\textwidth]{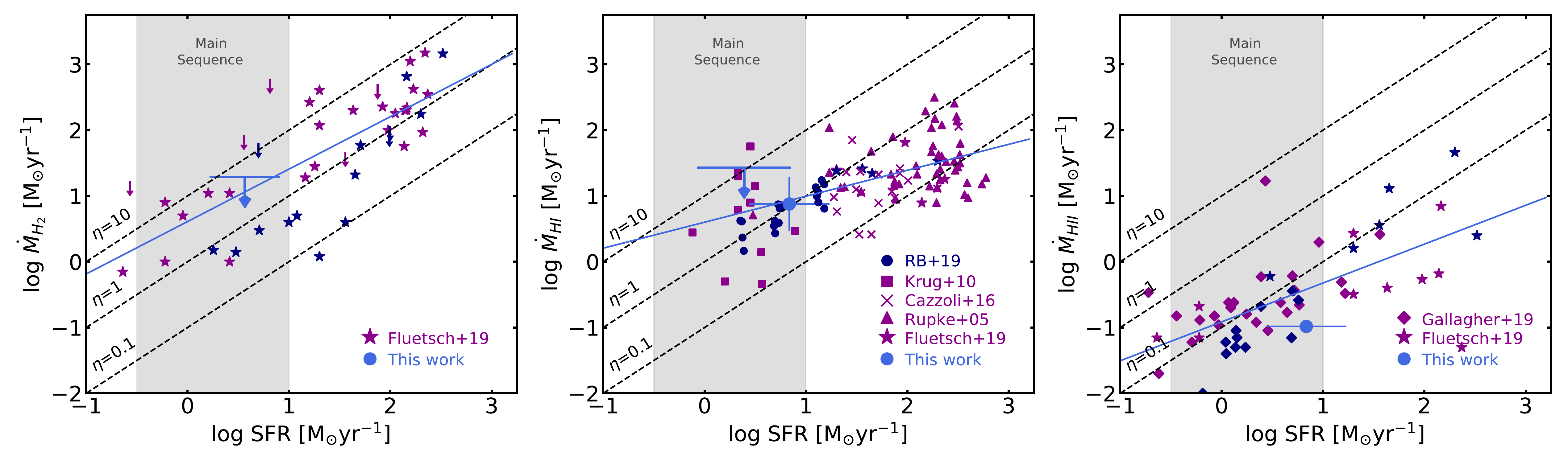}
 \caption{The mean H$_{2}$ (left), \HI\ (middle) and ionised (right) mass outflow rates compared to the mean SFRs of normal galaxies at $z\sim$0, as traced by CO(1-0), \HI\ 21cm and \nad, and H$\alpha$, respectively. Upper limits of the H$_{2}$ and \HI\ 21cm mass outflow rates are marked by blue arrow upper limits and the rates derived with \nad\ and H$\alpha$ are marked by blue circles with errors. The errors on the outflow rates quoted here include the 1$\sigma$ errors from our fitted parameters, whilst the error associated with the mean SFR is simply the standard deviation over all the galaxies in the stack (and galaxy spaxels, in the case of MaNGA data). The results from this study (light blue points and arrows) are compared to the mass outflow rates derived for \Hii\ (navy points) and (U)LIRG/AGN (purple points) galaxies from the literature, using multi-transition CO for molecular gas, \nad\ or C$^{+}$ for neutral gas, and  H$\alpha$ or H$\beta$ for ionised gas. The classification of galaxy type is taken from each relevant study. We further mark the region of SFRs that roughly probe the lower and upper limits of the high mass (M$_{*}\geqslant$10$^{10}$ M$_{\odot}$) galaxy MS (gray shaded region) and draw dashed lines to represent constant mass loading factors of $\eta$=0.1, 1, and 10. Light blue lines represent straight line fits to the full samples in each of our panels.}
 \label{fig:dmdtsfr}
\end{figure*}

Although we can derive only upper limits on the average, total mass outflow rate for outflows in normal galaxies in the local Universe, such constraints are extremely valuable. As such, in Figure \ref{fig:dmdtsfr} we plot our derived rates as a function of the mean SFR and compare these to several other results in the literature, namely those of \citet{rupke05b}, \citet{krug10}, \citet{cazzoli16}, \citet{fluetsch19}, \citet{gallagher19} and \citet{rb19}. In the case of \citet{krug10} and \citet{rb19}, we rederive the neutral gas outflow rates using Equation \ref{eq:finaldM} and their fitted parameters to \nad\ profiles to ensure similar assumptions are made in the conversion of $N$(\HI) column densities and outflow rates. The list of studies we compare to is not a complete one, however we are still able to provide useful comparisons to a variety of galaxy types and properties in the local Universe.

In comparing outflow rates of molecular gas in the left column of Figure \ref{fig:dmdtsfr}, we find our upper limit of $\dot{M}_{\text{CO}}<$19.43\,M$_{\odot}$yr$^{-1}$ consistent with the \Hii\ starburst and AGN galaxies from the compilation by \citet{fluetsch19} within the marked MS region, although $\sim$1-2 dex lower than the maximum rates found over their full sample. This is perhaps unsurprising given the nature of the galaxies and SFRs probed: most of the \citet{fluetsch19} sample are nuclear starbursts or strong AGN that reside above the galaxy MS with SFR$>$10\,M$_{\odot}$\,yr$^{-1}$ whilst we probe normal galaxies with a mean SFR less than 5\,M$_{\odot}$\,yr$^{-1}$. However, for the \Hii\ galaxies on the MS, their maximum and median mass outflow rate are 4\,M$_{\odot}$\,yr$^{-1}$ and 2.25\,M$_{\odot}$\,yr$^{-1}$, respectively, consistent with our upper limit which is a factor of $\sim$5$\times$ and $\sim$8.5$\times$ larger, respectively. These \Hii\ dominated galaxies are NGC 3628, NGC 253, NGC 1808, NGC 6810, and M82; all local, nuclear starburst galaxies which show evidence for a (sub)kpc outflow (with the exception of NGC 6810, for which no detection is found) in emission line maps of CO(1-0), except for NGC 253, whose outflow is characterised with maps of CO(2-1) emission (see Tables 1 and 2 of \citealt{fluetsch19}, and references therein). This consistency also extends to the AGN detections on the MS, which show maximum and median mass outflow rates of 11\,M$_{\odot}$\,yr$^{-1}$ and 6.5\,M$_{\odot}$\,yr$^{-1}$, respectively, not too dissimilar to our upper limit and suggestive of limited additional contribution from normal AGN to molecular outflows.

The comparison of an atomic mass outflow rate, traced by \HI\ 21cm emission, is somewhat challenging due to the large upper limit of $\dot{M}_{\text{HI}}<$26.72\,M$_{\odot}$yr$^{-1}$ set by this work and the scarcity of outflow rates derived directly from \HI\ emission in the literature. In the middle column of Figure \ref{fig:dmdtsfr}, we compare our rate to those derived for a variety of galaxy types through the indirect neutral gas tracers \nad\ and [\cii]. As with the molecular gas comparison, we find our $\dot{M}_{\text{HI}}$ to be consistent with the rates found over the full galaxy sample from the literature, whilst sitting $\sim$1 dex lower than the maximum value ($\sim$316 M$_{\odot}$yr$^{-1}$) found for (U)LIRGs by \citet{rupke05b}. For galaxies residing on the MS region only, our rate of $\dot{M}_{\text{HI}}<$26.72\,M$_{\odot}$yr$^{-1}$ is similar only to the maximum \nad-derived values of (U)LIRG/Seyfert objects from \citet{krug10} over the MS ($\sim$23-56 M$_{\odot}$yr$^{-1}$) and characterised by a mass loading factor of $\eta\lesssim$11. A comparison of the \HI\ outflow rate derived with \nad\ provides a more useful comparison, given the detection of blueshifted absorption which reveals a lower outflow rate compared to the upper limit from \HI\ emission: we find in general our derived value of 7.55\,M$_{\odot}$yr$^{-1}$ is consistent with other \nad\ outflow rates across the literature, as well as the relation and evolution with SFR observed by \citet{rb19} for SDSS galaxies. Although uncorrected for inclination, the median outflow rate across the full range of SFRs is $\sim$15.8 M$_{\odot}$yr$^{-1}$, and for galaxies residing on the MS only the median outflow rate drops to 4.66 M$_{\odot}$yr$^{-1}$ which is much more similar to the \nad\ outflow rate we derive here. A closer comparison to virtually identical samples of galaxies from observations with the SDSS \citep{rb19} shows a slightly lower median outflow rate of 4.09 M$_{\odot}$yr$^{-1}$. Such a difference is most likely attributed to the limited galaxy area covered by a 3$''$ fiber, as well as such rates not being corrected for inclination. Again, for normal galaxies across the MS, we find little difference between galaxies likely to host an AGN and those without, reinforcing the notion that \textit{normal} AGN feedback is unlikely to be much more significant. Our derived $\dot{M}_{\text{NaD}}$ of 7.55\,M$_{\odot}$\,yr$^{-1}$ is consistent with previously derived values for normal $z\sim0$ galaxies \citep{martin12,rubin14,rb19} but a factor of $\sim$8.5$\times$ smaller than our 21cm upper limit. We find that the \nad\ outflow rate is also comparable to the upper limit of CO, but a factor of $\sim$2.6$\times$ smaller, and this ratio is similar to the results of \citet{fluetsch19} for starburst galaxies, using \nad\ and C$^{+}$ as neutral gas tracers.

Finally, a comparison of the ionised gas outflow rate in the right column of Figure \ref{fig:dmdtsfr} further illustrates the similarity (or consistency) between our derived values and those in the studies we compare to (for both AGN and \Hii\ galaxies). The values in the MS region are again consistent with an evolution relating to the SFR and the median mass outflow rate of all \Hii\ (AGN) galaxies from the studies of \citet{fluetsch19} and \citet{gallagher19} in the MS region is 0.07\,M$_{\odot}$\,yr$^{-1}$ (0.2\,M$_{\odot}$\,yr$^{-1}$), virtually identical to our measured value of 0.10\,M$_{\odot}$\,yr$^{-1}$. In both this work and the study of \citet{fluetsch19}, the ionised outflow rate is by far the lowest of the three gas phases and essentially insignificant compared to the molecular and neutral gas phases.

Through our stacked points and upper limits, we have added constraints on mass outflow rates of different gas phases, and compared these to other results found for local galaxies, where general agreement and consistency is found. Using \nad\ as the tracer of neutral gas and summing the outflow rates over all the different gas phases, we find a total mass outflow rate of $\dot{M}_{\text{tot}}\lesssim$27\,M$_{\odot}$yr$^{-1}$ (or $\dot{M}_{\text{tot}}<$34\,M$_{\odot}$yr$^{-1}$, assuming the upper limits of our derived values in each gas phase), significantly higher than what is typically probed by tracers of a single gas phase. In comparing the different gas tracers (CO, \nad, H$\alpha$) to each other, we find the molecular and neutral gas phases have the potential to be by far the dominant outflowing phases, with $\sim$99\% of the total mass outflow rate coming from these two gas phases ($\lesssim$72\% from the molecular gas and $\sim$28\% from the neutral gas) and less than 1\% coming from the ionised gas. Between the molecular and neutral gas phases, the molecular gas has the potential to be by far the dominant gas phase, similar to findings by \citet{fluetsch19} for AGN. Such a picture is even more evident when comparing the data points in Figure \ref{fig:dmdtsfr} at SFR$\gtrsim$10 M$_{\odot}$yr$^{-1}$, where we see a significant decrease in outflow rates from the molecular to the ionised gas phase. The data presented here have Pearson correlation coefficients of 0.75, 0.53 and 0.67 for molecular, neutral and ionised gas, respectively. To quantify these differences further, we fit straight lines to the all the data points in each panel (using the
\nad\ constraints instead of \HI\ for the neutral gas), resulting in the following relations that can be used for estimates in simulations and observations:

\begin{align}
&\text{log}\,\dot{M}_{\text{H}_{\text{2}}} = 0.80\cdot\text{log SFR} + 0.61, \\
&\text{log}\,\dot{M}_{\text{HI}} = 0.39\cdot\text{log SFR} + 0.60, \\
&\text{log}\,\dot{M}_{\text{HII}} = 0.59\cdot\text{log SFR} - 0.91.
\end{align}

These relations are shown as light blue lines in Figure \ref{fig:dmdtsfr}. From the normalisation alone, the similarities between the molecular and neutral outflow rates over the full samples become apparent, while the ionised outflow rate appears vastly reduced in comparison. Additionally, by factoring in the slope of each relation, it becomes clear that the outflow rates of molecular gas most strongly correlate with the underlying SFRs of the galaxies, bringing forth important implications for the quenching potential of galaxies residing on and above the main sequence.

Thus, the emerging picture from our comparisons here are that galaxies of the local Universe display a large range of mass outflow rates, with the highest rates appearing in galaxies with higher SFRs. Galaxies residing on the MS display far reduced rates compared to their starburst counterparts and this applies across all phases. No significant difference is found between the outflow rates of \Hii\ galaxies and AGN, suggesting limited enhancement from AGN feedback \citep{sarzi16,rb19}. The molecular gas phase - the most important for star formation - likely contributes the majority of the total outflow rate, with the neutral gas phase contributing similar or slightly smaller fractions, and the ionised gas contributing negligible amounts. Specifically, using \nad\ as the tracer of neutral gas, we find relative fractions of $\dot{M}_{H_{2}}$/$\dot{M}_{\text{HI}}\lesssim$2.6 ($\dot{M}_{H_{2}}$/$\dot{M}_{\text{HI}}<$5.0 assuming upper limits and including errors) and $\dot{M}_{H_{2}}$/$\dot{M}_{\text{HII}}\lesssim$187 ($\dot{M}_{H_{2}}$/$\dot{M}_{\text{HII}}<$228 assuming upper limits and including errors).

At this point it is important to note that while our data sets, selection criteria and analyses ensure we conduct our search over representative samples of normal, star-forming galaxies, the MaNGA galaxies have specifically been chosen due to their association with blueshifted \nad\ absorption in the central regions, whilst the selection of xCOLD GASS and ALFALFA+xGASS samples are done blindly. This is purely due to the lack of \textit{a priori} constraints on the presence of outflows in individual objects over the latter two data sets, which is not the case for the MaNGA galaxies. Thus, the neutral and ionised mass outflow rates from the MaNGA data sets are inevitably better constrained and we note that were we to stack all of the galaxies in the \citet{rb20} parent sample, the resulting \nad\ profile would be in net emission. Given we can only report upper limits on the atomic and molecular mass outflow rates, however, such bias is unlikely to drive strong variations in the overall conclusions of the paper.

\subsection{Why Don't We See Atomic or Molecular Outflows?}
In Section \ref{sec:res} we have seen that clear detections of neutral and ionised outflows - as traced by \nad\ blueshifted absorption and broad H$\alpha$ emission - are seen in normal galaxies, but similar detections of molecular and atomic gas outflows, as traced by CO(1-0) and \HI\ 21cm emission, appear scarce. Several reasons exist to explain this. The first is that despite the high S/N of our stacked spectra, the mean spectra do not probe a low enough RMS necessary to observe outflowing emission in normal, star-forming galaxies. Given outflow signal in starburst and powerful AGN typically reach only $\sim$1-10\% of the maximum flux \citep{cicone14,ciconenature,fluetsch19}, lower fractions still might be expected for observations of less extreme objects. If we compare the depth of our CO and \HI\ spectra to the spectrum of the ionised gas, where broad emission is seen, we find the optical spectra a factor of $\sim$1.7$\times$ more sensitive than what our molecular and atomic gas stacks probe. In comparing to the data points of \citet{fluetsch19}, we find both our CO and their upper limits all lie close to the $\eta\sim$10 line, but virtually all of the detections lie below. Furthermore, for galaxies with SFR$\leqslant$10\,M$_{\odot}$yr$^{-1}$ (i.e., on the MS), all of the detections lie below a mass outflow rate threshold of $\sim$10\,M$_{\odot}$yr$^{-1}$ and non-detections above. For \Hii\ objects on the MS, we find a median upper limit (combining both our data set and the upper limit of \citealt{fluetsch19}) of $\lesssim$42\,M$_{\odot}$yr$^{-1}$, however the median outflow rate of \Hii\ galaxy detections is 2.25\,M$_{\odot}$yr$^{-1}$, more than a factor of $\sim$18$\times$ lower. At this point it is important to note that for each of the \citep{fluetsch19} starburst galaxies on the MS (NGC 3628, NGC 253, NGC 1808, NGC 6810, and M82) the molecular outflow was discovered and characterised via emission line maps of CO, rather than direct detections of broad outflow emission in integrated spectra. As such, it is quite likely that our stack simply does not probe the depth necessary to detect the smaller outflow rates of normal star-forming galaxies.

A second consideration is whether CO molecules and \HI\ atoms survive the turbulent and harsh environments of the outflow: shock fronts and UV radiation in strong outflows can accelerate and heat outflowing gas to high velocities and temperatures, leading to the photodissociation and ionisation of the gas. If this is the case, molecular and atomic gas that isn't properly shielded (either from self-shielding or from dust) could evaporate in the outflow. However, the contribution from shocks is thought to be relatively minor  \citep{cicone12,zubovas14} and, given the detection of CO outflows in starbust and AGN galaxies from a variety of sources in the literature, we deem this scenario unlikely.

Finally, a last consideration is that galaxy-wide outflows don't entrain atomic or molecular gas. Although this is clearly not the case for more extreme or exotic objects, the outflows observed in normal galaxies are found to be far less extreme, with much smaller velocities and outflow rates. As such, whether the low energetics of the outflow are able to disrupt the gravitationally-bound molecular gas in giant molecular clouds is subject to debate.

\subsection{Can Outflows Quench Normal Galaxies?}
\begin{figure}
\centering
 \includegraphics[width=0.85\columnwidth]{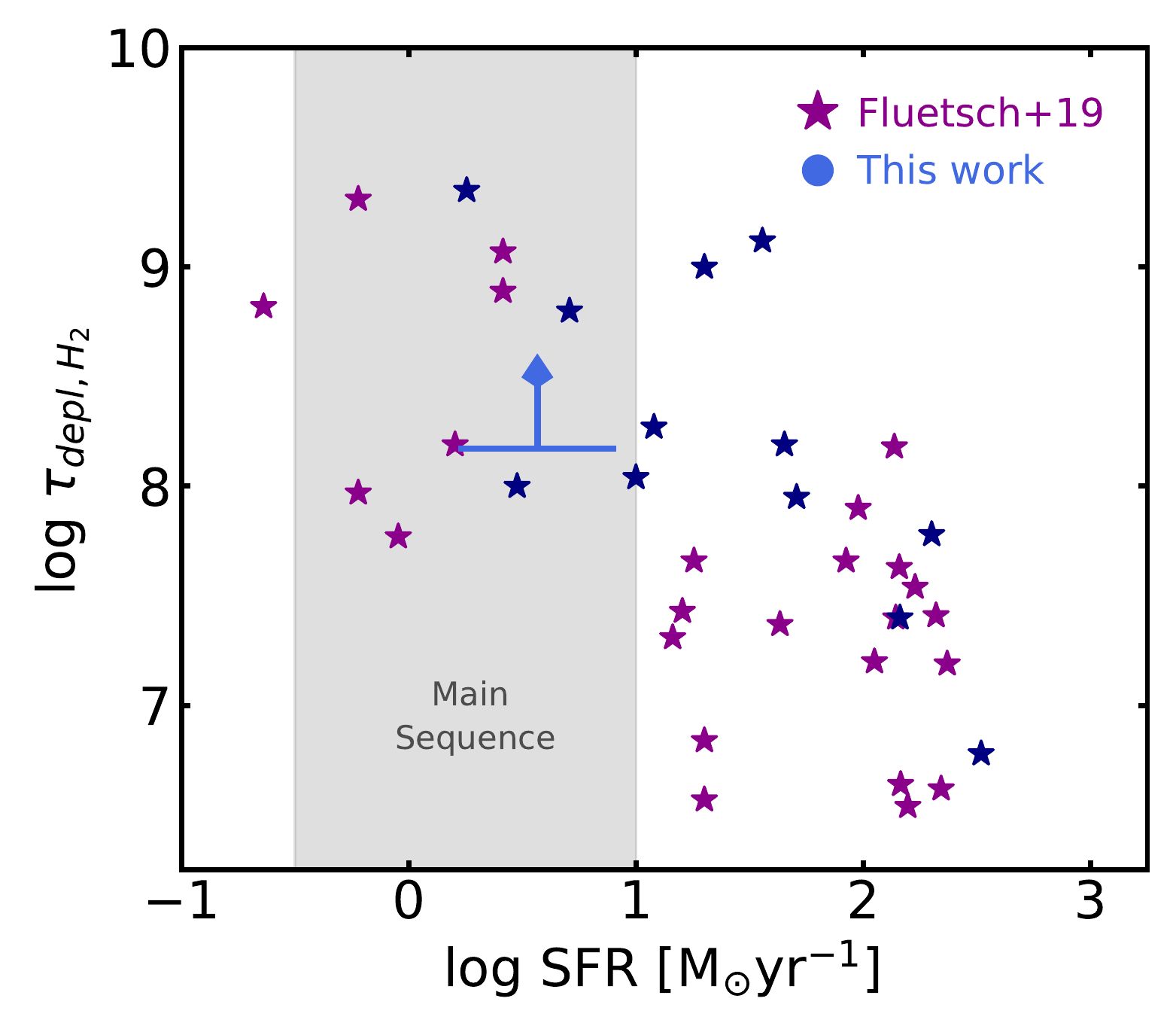}
 \caption{The outflow depletion time of molecular gas, defined as $\tau_{\text{depl}}$=$M_{\text{gal}}$(H$_{2}$)/$\dot{M}_{\text{out}}$(H$_{2}$), as a function of host SFR. The colour scheme is the same as in Figure \ref{fig:dmdtsfr} and we additionally plot the values found for starburst and AGN galaxies in \protect\citet{fluetsch19}.}
 \label{fig:tdepl}
\end{figure}

The crucial questions we aim to address in this study are: what are the relative fractions of outflowing gas from different gas phases and can outflows from normal galaxies quench the star formation in their hosts? In Section \ref{subsec:multiphasedmdt} we addressed the former. To constrain the latter, we look to the mass loading factors of our different gas phases. By dividing the mass outflow rates by the associated (total) SFRs, we find mass loading factors of $\eta\lesssim$5.27, $\eta\sim$1.10 and $\eta\sim$0.02 for molecular (CO), neutral (\nad) and ionised (H$\alpha$) gas, respectively. Considering only one gas phase, however, can severely underestimate the total mass loading factor due to reduced mass outflow rates, as seen in Section \ref{subsec:multiphasedmdt}. Thus, under the reasonable assumption that our three mass outflow rates are derived from similar distributions of galaxies with similar (average) SFRs (see Table \ref{tab:meanprops}), we can approximate the total, multiphase mass loading factor ($\eta_{\text{tot}}$=$\eta_{\text{CO}}$ + $\eta_{\text{NaD}}$ + $\eta_{\text{H}\alpha}$) and find a value of $\eta_{\text{tot}}\lesssim$6.39 (or $\eta_{\text{tot}}<$10.87 assuming upper limits and including errors). At face value, this would suggest that outflows are able to expel enough gas to eventually deplete the gas reservoirs of their host galaxies. However, this is largely reliant on the true value of molecular gas outflow rate lying close to our upper limit. If we compare our outflow rates for neutral and ionised gas only, we find a mass loading factor of $\eta_{\text{NaD}}+\eta_{H\alpha}\sim$1.12 (upper limit of $\eta <$2.58), a factor of $\sim$2.5$\times$ lower than our derived upper limit of the total mass loading factor and suggestive of the molecular gas playing a crucial role. As such, given the importance of the gas phase and the likely large contribution to the total mass outflow rate, in Figure \ref{fig:tdepl} we look at the outflow depletion time of molecular gas, $\tau_{\text{depl}}$=$M_{\text{gal}}$(H$_{2}$)/$\dot{M}_{\text{out}}$(H$_{2}$), which is the time required to completely remove all the molecular gas from the host assuming no additional gas is transferred to or from the galaxy.

Using a mean galaxy molecular gas mass of log\,M$_{H_{2}}\sim$9.46 as probed by our CO stack, we derive a lower limit depletion time of $\gtrsim$150 Myr (or $\gtrsim$10$^{8.17}$ years), suggestive of a potentially rapid quenching process. This is consistent and similar to the depletion times found by \citet{fluetsch19} for \Hii\ and AGN galaxies on the MS, but significantly longer than values found for strong starbursts and AGN which, whilst rarer, are able to remove significantly larger amounts of gas on shorter timescales.

However, such a comparison is only valid to first order, and the question of whether outflows can quench their host galaxies or not via ejective feedback is significantly more complex: important to consider are also the outflow velocities relative to the escape velocity of the galaxy, whether the host galaxy has its gas reservoir replenished at any point, and whether the fate of an outflow lies in clearing out into the CGM or falling back to the galaxy disk in the shape of a galactic fountain. Although our data set does not allow us to directly confirm such a scenario, our derived outflow velocities and mass loss rates suggest that strong ejective feedback in normal star-forming galaxies is absent and the relatively weak outflows seen here are unlikely to eject enough gas completely out of the galaxy disk and into the CGM to induce full galaxy-wide quenching, thereby moving it from the MS to the red-sequence population.

\section{Summary \& Conclusions}
We have used observations of CO(1-0), \HI\ 21cm, \nad, and H$\alpha$ from the xCOLD GASS, xGASS, ALFALFA $\alpha$.100 and MaNGA DR15 surveys to constrain the prevalence and relative fractions of molecular, neutral and ionised outflows in normal galaxies in the local Universe. Using stacking techniques to create high S/N composite spectra, we have performed extensive tests to ensure broad emission from outflows is unlikely to be caused by confusion, artefacts or RFI and stack 65 CO(1-0), 1,077 \HI\ and 78 optical spectra. Our findings based on these stacked spectra are the following:

\begin{itemize}
\item Outflowing gas is observed in both \nad\ and H$\alpha$ tracers through significant blueshifted absorption and broad emission, respectively. From these tracers we infer inclination-corrected outflow velocities of $|$v$_{\text{out}}|$=160 \kms\ and $|$v$_{\text{out}}|$=534 \kms\ and mass outflow rates of $\dot{M}_{\text{NaD}}$=7.55$\pm$7.20 M$_{\odot}$\,yr$^{-1}$ and $\dot{M}_{H\alpha}$=0.10$\pm$0.02 M$_{\odot}$\,yr$^{-1}$, respectively.
\\
\item In contrast to the optical tracers, we observe no significant broad emission on either side of the CO(1-0) or \HI\ 21cm emission - although do observe tentative broadening in the \HI\ stacks - despite the high sensitivity of our stacks. As such, we place upper limits on the molecular and atomic mass outflow rates of $\dot{M}_{\text{CO}}<$19.43\,M$_{\odot}$yr$^{-1}$ and $\dot{M}_{\text{HI}}<$26.72\,M$_{\odot}$yr$^{-1}$, based on model completeness and reliability at the S/N of our stacks and assuming an outflow velocity of 200 \kms.
\\
\item Combining the upper limits of the molecular (CO(1-0)) outflow rate with our detections of outflowing neutral (\nad) and ionised (H$\alpha$) gas, we derive a total mass outflow rate of $\dot{M}_{\text{tot}}\lesssim$27\,M$_{\odot}$yr$^{-1}$, with gas phase ratios of $\dot{M}_{\text{CO}}$/$\dot{M}_{\text{NaD}}\lesssim$2.6 and $\dot{M}_{\text{CO}}$/$\dot{M}_{H\alpha}\lesssim$187. Dividing each mass outflow rate by the associated galaxy-wide SFR and summing the ratios, we find an approximate total, multiphase mass loading factor of $\eta\lesssim$6.39 ($\eta<$10.87 including errors), suggesting that estimates of single gas-phase outflow rates are likely to miss significant amounts of outflowing gas. The derived upper limit on the total mass loading factor suggests that, to first order, outflows from normal outflows may cause some degree of localised quenching in the host galaxies. However, accurate determination of the degree of quenching requires careful consideration of the low outflow velocities and fate of the gas, hydrodynamical deceleration by surrounding gas and disk gravity, and knowledge of gas inflow rates used to replenish the cold gas reservoirs.
\end{itemize}

The multiphase nature of outflows is a crucial constraint towards our understanding of how outflows impact their host galaxies, as it allows us to estimate the total gas mass being ejected out of the galaxy, as well as the relative importance of the difference gas phases. Tracers of the molecular gas phase are by far the most valuable, since molecular gas directly impacts the star formation in the disk. However, such constraints are hard to come by and have for the most part only been placed on extreme or nuclear starbursts or AGN via emission line mapping. Here we have progressed on this deficiency by placing important constraints on the multiphase nature of outflows over normal galaxy populations at $z\sim$0 with stacking methods of large samples of galaxies. However, despite the high S/N of our stacks, molecular and atomic (\HI) gas outflows remain undetected. As such, to gain a better understanding of the prevalence of molecular and atomic gas outflows - as well as their quenching potential - a dedicated observational program over a representative sample of galaxies with e.g., ALMA is necessary to reach the required depths in a search for molecular and atomic outflow signatures in normal galaxies.
Finally, with a view to gaining a complete understanding of the outflows' impact on the galaxy system, still required is a thorough understanding of the fate of the expelled gas, and whether the impact of outflows on galaxies is exclusive to ejective feedback or whether preventive feedback can play an important role in halting accretion to the host galaxy. Such dedicated observations, in combination with accurate simulations of outflows and the CGM, would greatly contribute to our understanding of outflows and ultimately their potential for converting star-forming galaxies on the MS to passive, ``red and dead'' galaxies where star formation has been virtually extinguished.

\section*{Acknowledgements}
GRB would also like to acknowledge and extend his sincere thanks to Am\'elie Saintonge for her generous and valuable contributions and feedback to this work. Many thanks are also due to Barbara Catinella, for useful discussions on criteria for \HI\ confusion, Martha Haynes for providing the ALFALFA $\alpha$.100 spectra, and Thomas Greve and Marc Sarzi for discussions that improved this paper. GRB would also like to thank the anonymous referee for their helpful comments and suggestions. This research was supported by grants from the Royal Society. \\
Funding for the Sloan Digital Sky Survey IV has been provided by the Alfred P. Sloan Foundation, the U.S. Department of Energy Office of Science, and the Participating Institutions. SDSS acknowledges support and resources from the Center for High-Performance Computing at the University of Utah. The SDSS web site is www.sdss.org. \\
SDSS is managed by the Astrophysical Research Consortium for the Participating Institutions of the SDSS Collaboration including the Brazilian Participation Group, the Carnegie Institution for Science, Carnegie Mellon University, the Chilean Participation Group, the French Participation Group, Harvard-Smithsonian Center for Astrophysics, Instituto de Astrofísica de Canarias, The Johns Hopkins University, Kavli Institute for the Physics and Mathematics of the Universe (IPMU) / University of Tokyo, the Korean Participation Group, Lawrence Berkeley National Laboratory, Leibniz Institut für Astrophysik Potsdam (AIP), Max-Planck-Institut für Astronomie (MPIA Heidelberg), Max-Planck-Institut für Astrophysik (MPA Garching), Max-Planck-Institut für Extraterrestrische Physik (MPE), National Astronomical Observatories of China, New Mexico State University, New York University, University of Notre Dame, Observatório Nacional / MCTI, The Ohio State University, Pennsylvania State University, Shanghai Astronomical Observatory, United Kingdom Participation Group, Universidad Nacional Autónoma de México, University of Arizona, University of Colorado Boulder, University of Oxford, University of Portsmouth, University of Utah, University of Virginia, University of Washington, University of Wisconsin, Vanderbilt University, and Yale University.


%%%%%%%%%%%%%%%%% APPENDICES %%%%%%%%%%%%%%%%%%%%%
\appendix
\section{Optical Images and Radio Spectra of Galaxies}
\label{sec:appendixa}
Here we present plots of the SDSS images, xCOLD GASS CO(1-0) spectra and xGASS/ALFALFA \HI\ spectra of a random selection of 10 galaxies from the sample described in Section \ref{sec:subsamps_sec}.

\begin{figure*}
\centering
 \includegraphics[width=0.725\textwidth]{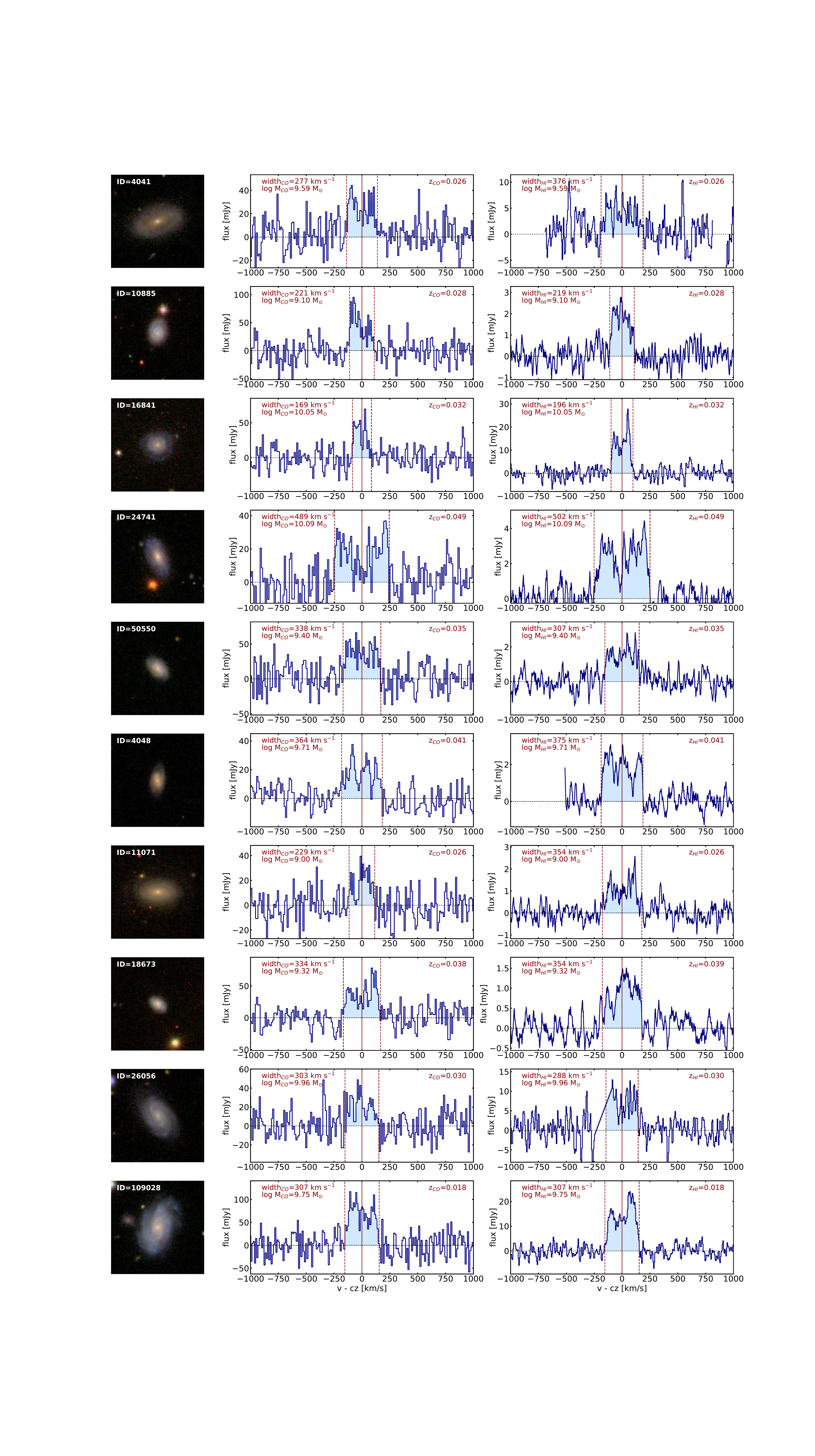}
 \caption{The 1.25$'$x1.25$'$ SDSS image (left), IRAM 30m CO(1-0) spectrum (middle) and Arecibo \HI\ 21 cm spectrum (right) for 10 randomly-selected galaxies from the samples defined in Section \ref{sec:subsamps_sec}. Each spectrum has been checked for confusion (within a normalised velocity of $\pm$5) and baseline issues and the quoted widths and velocities are those measured in the study.}
 \label{fig:xCGpostage}
\end{figure*}

%%%%%%%%%%%%%%%%%%%%%%%%%%%%%%%%%%%%%%%%%%%%%%%%%%

% Don't change these lines
\bsp  % typesetting comment
\label{lastpage}
\end{document}